\documentclass[12pt]{article}

\usepackage{graphicx}
\usepackage{qcmc06}

\begin{document}

\pagestyle{empty}

\title{Phase Sensitivity of a Mach-Zehnder Quantum Sensor}

\author{Thomas B. Bahder\refnote{1} and Paul A. Lopata\refnote{2}}

\affiliation{%
\affnote{1}Charles M. Bowden Research Facility, AMSRD-AMR-WS\\
Aviation and Missile Research, Development, and Engineering Center\\ 
US Army RDECOM, Redstone Arsenal, AL 35898 USA \\
\affnote{2}Army Research Laboratory\\ 2800 Powder Mill Road, Adelphi, MD 20783 USA}

\begin{abstract}
We investigate the dependence of the fidelity of a Mach-Zehnder quantum interferometer on the prior information about the phase, 
for Fock state input and for maximally entangled (N00N) state input. For no prior information, the fidelity for Fock state input is greater than for N00N state input.   In the limit of a narrow distribution describing the prior information, we find that both Fock and N00N state inputs lead to nearly equal fidelity.  
\end{abstract}

\setcounter{section}{0}

\section{Introduction}

The roots of quantum metrology has a history that dates back to Heisenberg's work in the early years of the previous century~\cite{Heisenberg1927}.  
More recently, the phase sensitivity of interferometers is of interest because quantum interferometers may be used as sensors of classical fields~\cite{Bahder2006}.   Fluctuations limit the precision of phase measurements~\cite{Caves1981} and the phase sensitivity of interferometers has been explored for different types of input states, such as squeezed states and number states, see references contained in~\cite{Bahder2006}.  Recently, there has been experimental progress in making efficient single-photon and entangled-photon sources, and photon number resolving detectors. Therefore, we concentrate on estimating the phase (and thereby the classical field) from photon number counts in the output ports of an  interferometer. 

The phase sensitivity $\Delta\phi$ is often discussed according to how it scales in two limits, known as the standard limit (shot noise), $\Delta \phi_{SL}=$ $1/\sqrt{N}$, and the Heisenberg limit, 
$\Delta\phi_{HL}=$ $1/N$, where $N$ is the number of particles that enter the
interferometer during each measurement cycle.  These results are based on
standard estimation theory\cite{Helstrom1976} which connects a continuous valued measurement outcome, $m$, with a corresponding phase, $\phi$, through a theoretical relation $m=m(\phi)$. The derivation of the Heisenberg limit of phase sensitivity implicitly assumes that the phase probability distribution has a single peak.   Using a Bayesian analysis we have shown that this assumption is not generally true~\cite{Bahder2006}, unless we know some prior information about the phase, for example that the phase is in some small region.  We have shown that the phase probability distributions have two peaks when Fock states are input into a Mach-Zehnder (MZ) interferometer and they have four peaks when maximally entangled states (N00N) are input~\cite{Bahder2006},  when there is no prior information about the phase.   Consequently, it is not clear whether the phase sensitivity, described in terms of the width of the phase probability distribution for measurement $m$, $\Delta \phi$, should refer to the width of a single peak or to the width of the whole distribution (containing all the peaks). As a result of this ambiguity, we have proposed a new metric of interferometer phase sensitivity~\cite{Bahder2006}, 
\begin{equation}
H(\Phi \colon \! M)= \sum_{m}\int_{-\pi}^{+\pi}d\,\phi\,\,P(m|\phi
) \,  p(\phi) \,\,\log_{2}\left[  \frac{ P(m|\phi)\,}{\int_{-\pi}^{+\pi}%
\,\,\,P(m|\phi^{\prime})  \, p(\phi^\prime) \,\,d\,\phi^{\prime}}\right]
\label{MutualInformationEq}%
\end{equation}
which we call the fidelity. This is the Shannon mutual information between the phase (a continuous alphabet) random variable, $\Phi$,  and the photon-number measurement outcomes (a discrete alphabet) with random variable $M$. Here  $P(m|\phi)$ is the conditional probability of a measurement outcome $m$  given the phase was $\phi$, and $p(\phi)$ is the {\it a priori} probability distribution that characterizes our prior knowledge of the phase. The fidelity $H(\Phi \colon \! M)$ is a statistical characterization of the phase sensitivity of the interferometer that is valid when multiple peaks exist in $p(\phi | m)$. The fidelity is a measure of the quality of a  sensor for given input state, and  it applies to both classical and quantum interferometers.  For a standard quantum MZ interferometer with a phase shift $\phi$ in one arm~\cite{Bahder2006}, with photon input into ports ``a" and ``b", the measurement outcomes are  $m=(n_c,n_d)$, where $n_c$ and $n_d$ are the photon numbers output in port ``c" and ``d".  The phase probability distribution is obtained from Bayes' rule as
\begin{equation}
p(\phi | m)=\frac{ P(m | \phi) \, p(\phi)}{\int_{-\pi}^{+\pi}P(m | \phi^{\prime}) \, p(\phi^\prime) \, \, d\phi^{\prime}},  
\label{PhaseProbabiliyDensity}%
\end{equation}
where \mbox{$-\pi < \phi \le \pi$}.   
In this report, we investigate the dependence of the fidelity of a MZ interferometer on the {\it a priori} information available, $p(\phi)$ , comparing Fock state and N00N state input. 
\begin{figure} 
\begin{minipage}[t]{18pc}
\includegraphics[width=18pc]{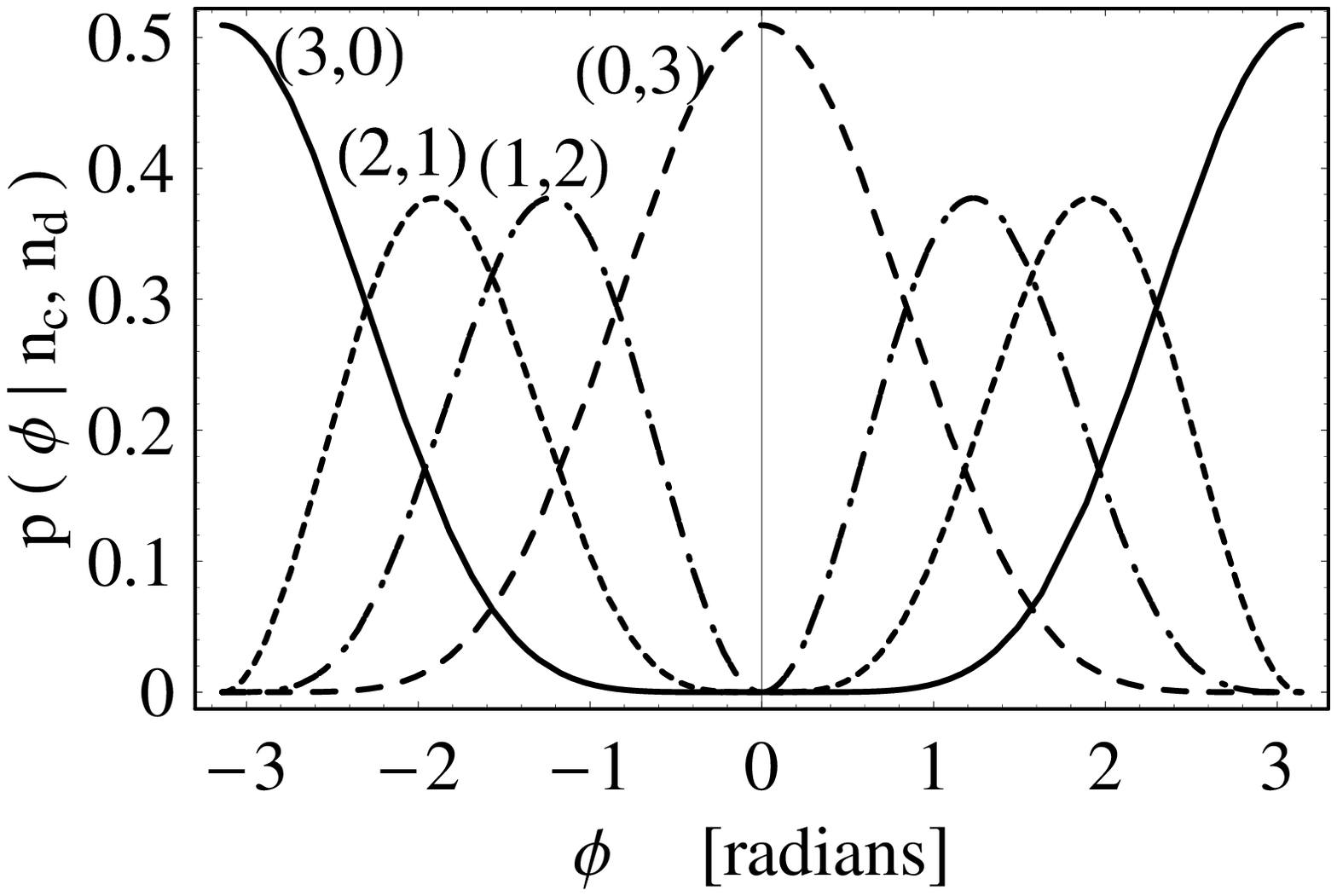}
\caption{\label{FIG1} The four phase probability distribution functions are shown for the measurement outcomes $(n_c,n_d)=(3,0)$, $(2,1)$, $(1,2)$, and $(0,3)$, for 3-photon Fock state input into port ``a" and vacuum input into port ``b", $|3,0\rangle$, for $p(\phi)=1/(2 \pi)$.}
\end{minipage}\hspace{2pc}%
\begin{minipage}[t]{18pc}
\includegraphics[width=18pc]{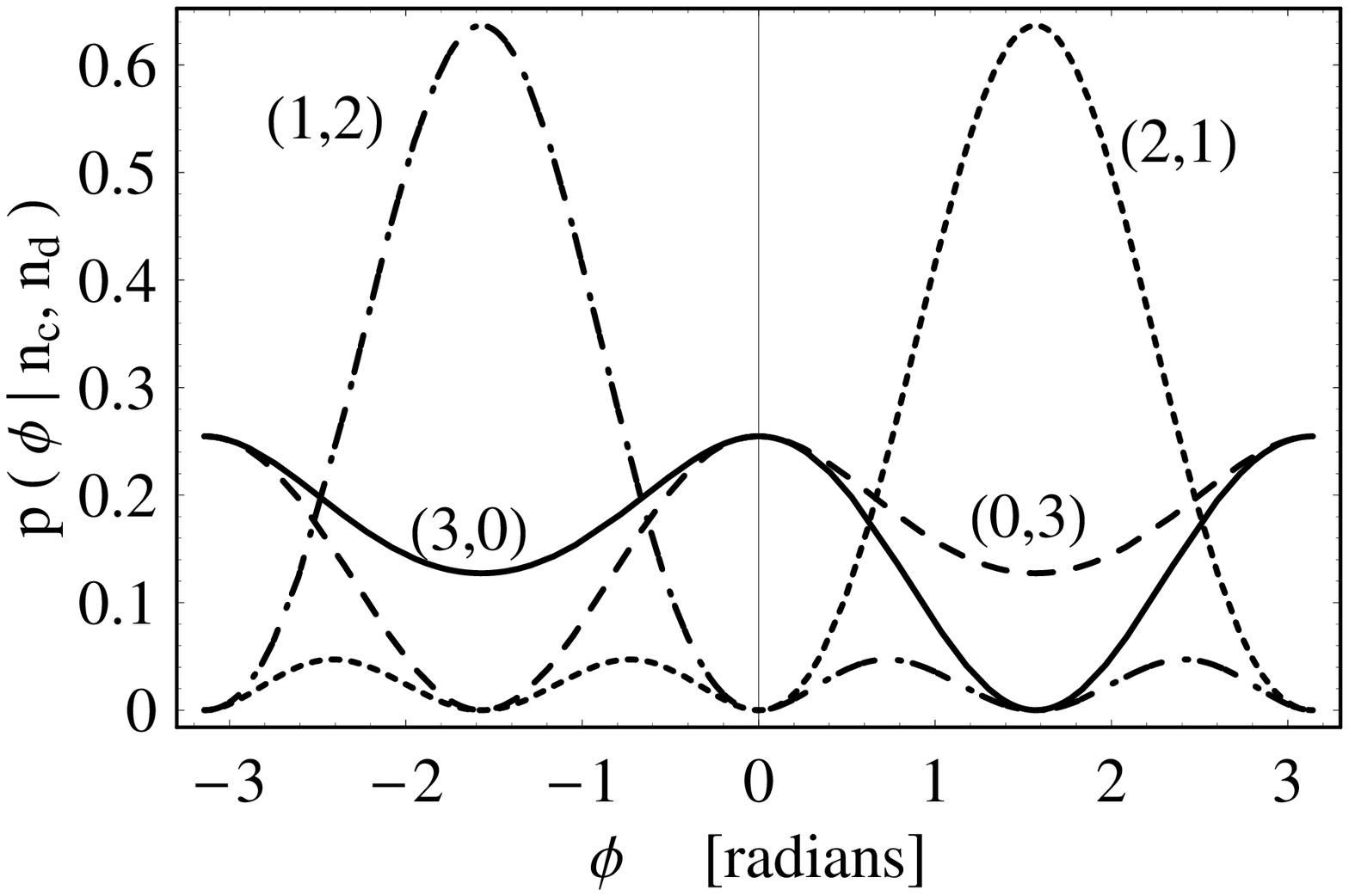}
\caption{\label{FIG1} The four phase probability distribution functions are shown for the measurement outcomes $(n_c,n_d)=(3,0)$, $(2,1)$, $(1,2)$, and $(0,3)$, for 3-photon N00N state input, $|3, 0 \rangle + |0, 3 \rangle$, for $p(\phi)=1/(2 \pi)$.}
\end{minipage} 
\end{figure}
The choice of input states is not intended to be optimum, but simply to compare two possible input states that are of current interest.

\section{Phase Sensitivity}

For N-photon Fock state input into arm  ``a" and vacuum input into port ``b", there are generally two peaks in the phase probability distribution $p(\phi | n_c, n_d)$, when $p(\phi)=1/(2 \pi)$.  For the same $p(\phi)=1/(2 \pi)$ with N-photon  N00N (maximally entangled) state input, $|N, 0\rangle + |0, N\rangle$, there are generally four peaks in $p(\phi | n_c, n_d)$~\cite{Bahder2006}.  For a single measurement, when a given measurement outcome is obtained, Eq.(\ref{PhaseProbabiliyDensity}) gives the probability distribution for the phase $\phi$.  

When a 1-photon Fock state is input into port ``a", $| 1, 0\rangle$, the probabilities for photon output into ports ``c" and ``d" are $\sin^2(\phi/2)$ and $\cos^2(\phi/2)$, respectively.  When a 1-photon N00N state is input, $(|1,0\rangle +|0,1\rangle)$, the probabilities for photon output in ports ``c" and ``d" are $\left[ \cos(\phi/2) - \sin(\phi/2)\right]^2 /2 $ and $\left[ \cos(\phi/2)+ \sin(\phi/2) \right]^2 / 2$,  respectively.  
(It is curious to note that these probabilities, in this extreme quantum limit of 1-photon input, are equal to the classical probabilities for input of energy into only port ``a", and input of equal energy into both ports ``a" and ``b" (classical analog of N00N state input)).   

For 2-photon input, there are three possible measurement outcomes. For 2-photon Fock state input into port ``a" and vacuum in port ``b", there are three probability density functions $p(\phi | m)$  for the phase:  $\frac{4}{3 \pi}  \sin^4 (\phi/2)$,   $\frac{1}{ \pi}  \sin^2 (\phi)$,   and $\frac{4}{3 \pi}  \cos^4 (\phi/2)$, for the measurement outcomes $m=(2_c,0_d) $, $(1_c,1_d)$, and $(0_c,2_d)$, respectively, for $p(\phi)=1/(2 \pi)$.     However, for the 2-photon N00N state input, the MZ interferometer behaves as a beam splitter, where there is 1/2 probability that both photons come out port ``c", 1/2 probability that both photons come out port ``d",  and zero probability that one photon comes out each port.   For this case, the probability of a given measurement outcome is independent of $\phi$. Consequently, the 2-photon N00N state cannot be used for determining the phase.  This is not the case with higher photon-number N00N state input.  

In Figures 1 and 2, we show the phase probability distributions for $N=3$ photon Fock state and N00N state inputs, respectively, for $p(\phi)=1/(2 \pi)$.  Note that for N00N state input,  the phase probability distributions are not symmetrical about $\phi=0$ and do not have simple peaks.
\begin{figure} 
\begin{minipage}[t]{18pc}
\includegraphics[width=18pc]{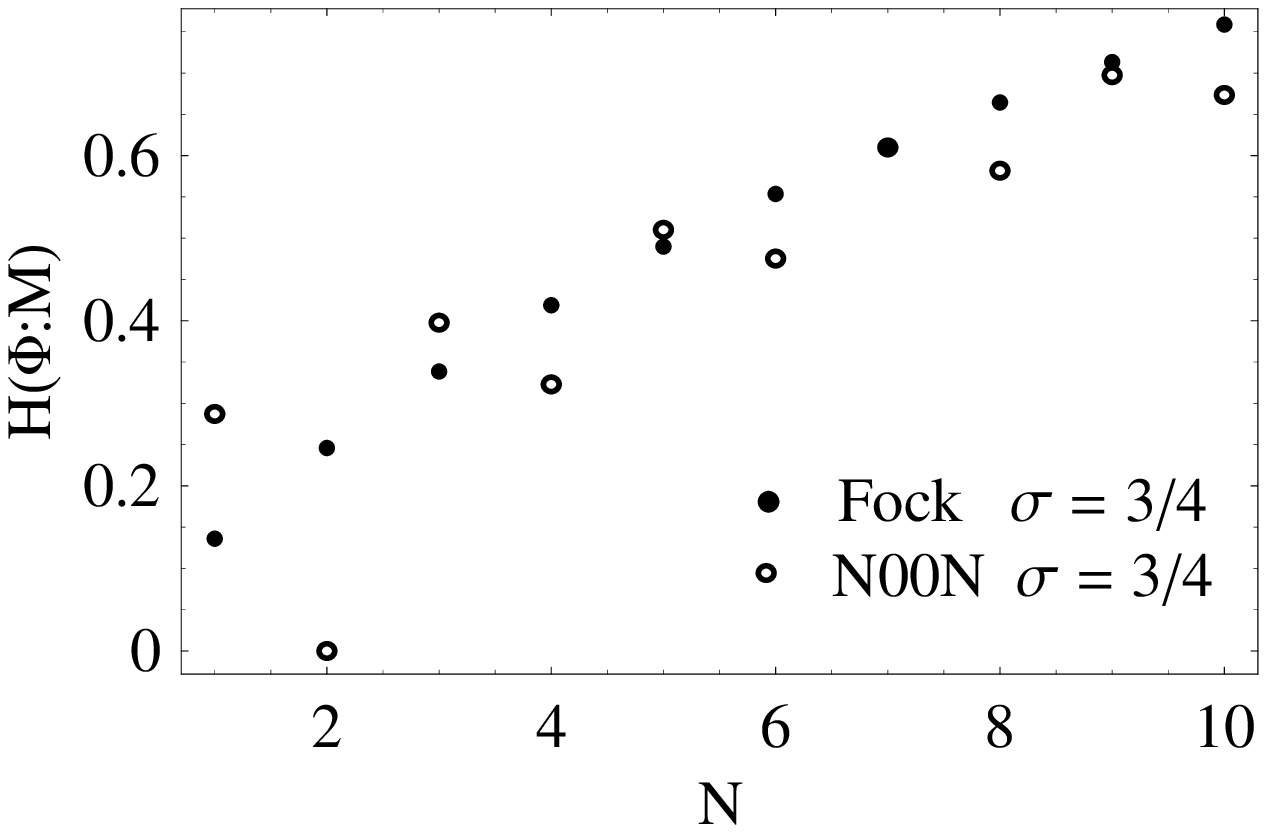}
\caption{\label{FIG1} The fidelity for Fock state and N00N state input with $p(\phi)$  given by a Gaussian centered at $\phi_0 = 0$ with $\sigma =3/4$.}
\end{minipage}\hspace{2pc}%
\begin{minipage}[t]{18pc}
\includegraphics[width=18pc]{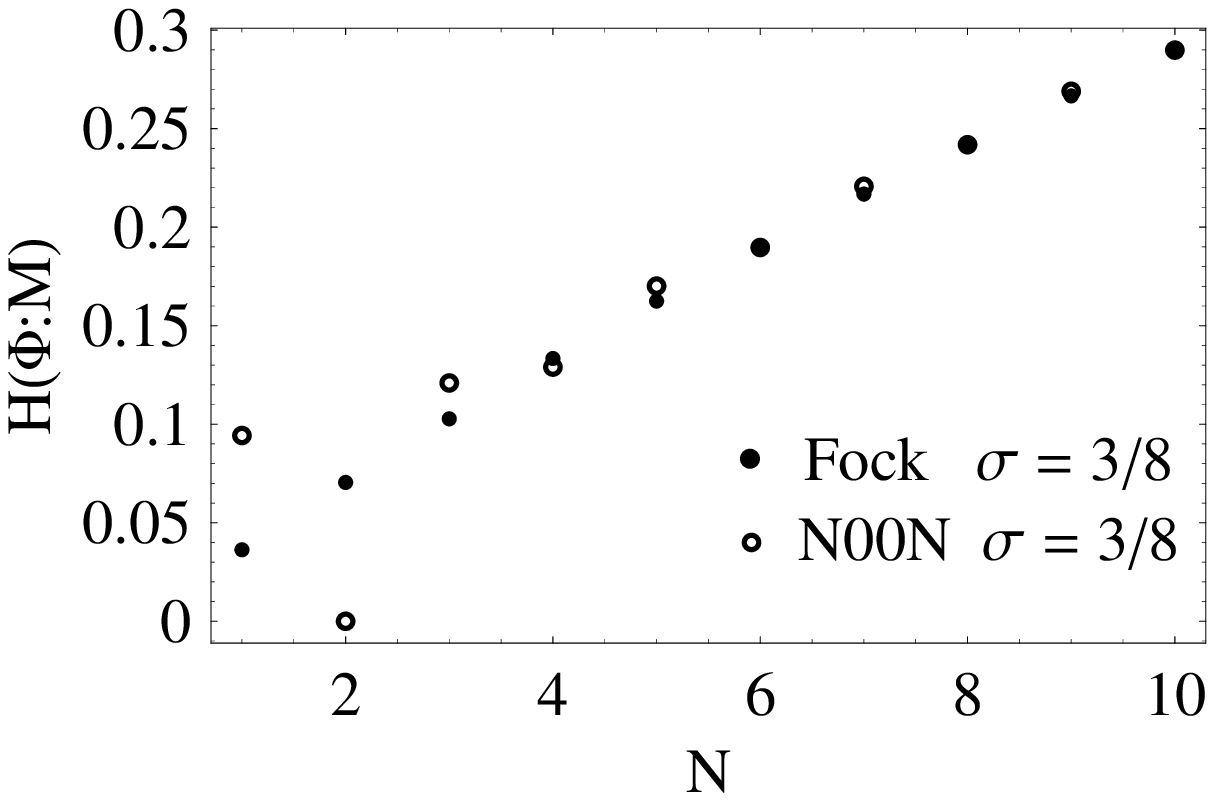}
\caption{\label{FIG1}  The fidelity for Fock state and N00N state input with $p(\phi)$  given by a Gaussian centered at $\phi_0 = 0$ with $\sigma =3/8$.}
\end{minipage} 
\end{figure}
As the number of input photons $N$ increase, the phase probability distributions get narrower.  A Gaussian fit to one peak of the probability distributions for Fock state inputs of $N=1$ to $N=40$ photons indicates that the phase sensitivity is given by $\Delta \phi_{Fock} = 0.897 / N^{0.477} $,  where $N$ is the number of photons input into port ``a".  However, the N00N state phase probability distributions do not have localized peaks, see Figure 2, and hence it is not possible to estimate the phase sensitivity from such widths. 

Consequently, we use the fidelity of the interferometer, given by Eq.~(\ref{MutualInformationEq}), to compare the statistical phase sensitivity on average, for Fock state input and for N00N state input. In previous work~\cite{Bahder2006}, we computed the fidelity of a MZ interferometer for the case of no prior information (taking $p(\phi)=1/(2 \pi)$) for Fock state input and for N00N state input, and we found that the fidelity for Fock state input is greater than or equal to that of N00N state input, for all input photon numbers~\cite{Bahder2006}.  

In the opposite regime, where we have very precise prior information, so $p(\phi)$ has one narrow peak about some value $\phi=\phi_0$, the fidelity can be approximated as
\begin{equation}
H(\Phi \colon M) = \frac{{\sigma ^2 }}{{2\ln 2}}\,\sum\limits_m {\left[ {P''(m|\phi _0 )\,\left( {1 - \ln 2} \right) + \frac{{\left[ {P'(m|\phi _0 )} \right]^2 }}{{P(m|\phi _0 )}}} \right]} 
\label{ShannonTightAPriori}  
\end{equation}
where $ \phi _0  = \int\limits_{ - \pi }^{ + \pi } {\phi \,p(\phi )} \,d\phi $, 
$\sigma ^2  = \int\limits_{ - \pi }^{ + \pi } {(\phi  - \phi _0 )^2 \,p(\phi )} \,d\phi  $, $P^{(n)}(m|\phi _0 ) = \left[ d^n P(m | \phi) / d \phi^n \right]_{\phi=\phi_0}$,  and where we assumed that $P(m | \phi_0)\ne 0$.  In the limit of narrow distribution $p(\phi)$, we find that the fidelity for Fock state input and N00N state input is equal, $ H(\Phi \colon M) = \sigma^2 N /(2\ln 2) $, 
for given total number of input photons $N$.  (The exception to this statement is for N00N state input, with $n_c = n_d$, when $n_c$ is odd, because then $ P(m |\phi ) = 0$.)  From these results we see that in both limiting cases, where we have no prior information, and in the limit of narrow distribution for the prior information, the fidelities are equal, so the use of N00N states does not provide an advantage over Fock states.  

Finally, we compare the fidelity in the intermediate regime, for Fock state and N00N state input, for the case when the prior information is given by a Gaussian function $ p(\phi ) = C(\phi_0 ,\sigma )\,\exp [ - \frac{{(\phi  - \phi_0)^2 }}{{2\sigma ^2 }}]
$,  where the normalization $C(\phi_0 ,\sigma )$ is given by  $ \int\limits_{ - \pi }^{ + \pi } {p(\phi )}  = 1$.  See Figures 3 and 4. While there are small variations in fidelity  for small photon numbers (notably for $N=1$ and $N=2$), we find that generally Fock state input produces a higher fidelity than N00N state input. In Figures 3 and 4, we have chosen a prior information represented by a Gaussian centered at $\phi_0=0$. The result for  $\phi_0  \not= 0$ must be still be determined.   Finally, we note that in actual applications the prior information may have a complicated functional form with multiple maxima.


\begin{thebibliography}{99}

\bibitem {Heisenberg1927}W. Heisenberg, Z. Phys. \textbf{43}, 172 (1927).

\bibitem{Bahder2006}  Thomas B. Bahder and Paul A. Lopata,  \textquotedblleft Fidelity of
Quantum Interferometers\textquotedblright , http://arxiv.org/abs/quant-ph/0602123, and
Phys. Rev. A \textbf{74}, 051801R (2006).  

\bibitem{Caves1981}C. M. Caves, Phys. Rev. D \textbf{23}, 1693 (1981). 

\bibitem{Helstrom1976}C.~W.~Helstrom, {\it Quantum Detection and Estimation Theory}, 
(Academic Press, New York, 1976).

\end{thebibliography}
\end{document}